\documentclass[pdflatex,sn-nature]{sn-jnl}% Math and Physical Sciences Numbered Reference Style
\usepackage{graphicx}%
\usepackage{multirow}%
\usepackage{amsmath,amssymb,amsfonts}%
\usepackage{amsthm}%
\usepackage{mathrsfs}%
\usepackage[title]{appendix}%
\usepackage{xcolor}%
\usepackage{textcomp}%
\usepackage{manyfoot}%
\usepackage{booktabs}%
\usepackage{algorithm}%
\usepackage{algorithmicx}%
\usepackage{algpseudocode}%
\usepackage{listings}%

\usepackage{makecell} % 必须，用于处理表头换行
\usepackage{float}
\theoremstyle{thmstyleone}%
\theoremstyle{thmstyletwo}%

\theoremstyle{thmstylethree}%

\begin{document}
\title{
Discovery of Interpretable Physical Laws in Materials via Language-Model-Guided Symbolic Regression
}

%%=============================================================%%

\author[1,2]{\fnm{Yifeng} \sur{Guan}}

\author[1,2]{\fnm{Chuyi} \sur{Liu}}

\author[1]{\fnm{Dongzhan} \sur{Zhou}}

\author[1]{\fnm{Lei} \sur{Bai}}

\author[3,4]{\fnm{Wan-jian} \sur{Yin}}

\author*[2,5]{\fnm{Jingyuan} \sur{Li}}\email{jingyuanli@zju.edu.cn}

\author*[1,6]{\fnm{Mao} \sur{Su}}\email{sumao@pjlab.org.cn}

\affil[1]{\orgname{Shanghai Artificial Intelligence Laboratory}, \orgaddress{\city{Shanghai}, \postcode{200232}, \country{China}}}

\affil[2]{\orgdiv{School of Physics}, \orgname{Zhejiang University}, \orgaddress{\city{Hangzhou}, \postcode{310058}, \country{China}}}

\affil[3]{\orgdiv{Soochow Institute for Energy and Materials Innovations (SIEMIS)}, \orgname{Soochow University}, \orgaddress{\city{Suzhou}, \postcode{215006}, \country{China}}}

\affil[4]{\orgname{Hefei National Laboratory},\orgaddress{\city{Hefei},\postcode{230088},\country{China}}}

\affil[5]{\orgdiv{Institute for Advanced Study in Physics}, \orgname{Zhejiang University}, \orgaddress{\city{Hangzhou}, \postcode{310058}, \country{China}}}

\affil[6]{\orgdiv{Shenzhen Institute of Advanced Technology}, \orgname{Chinese Academy of Sciences}, \orgaddress{\city{Shenzhen}, \postcode{518055}, \country{China}}}

%%==================================%%
%% Sample for unstructured abstract %%
%%==================================%%

\abstract{Discovering interpretable physical laws from high-dimensional data is a fundamental challenge in scientific research. Traditional methods, such as symbolic regression, often produce complex, unphysical formulas when searching a vast space of possible forms. We introduce a framework that guides the search process by leveraging the embedded scientific knowledge of large language models, enabling efficient identification of physical laws in the data. We validate our approach by modeling key properties of perovskite materials. 
Our method mitigates the combinatorial explosion commonly encountered in traditional symbolic regression, reducing the effective search space by a factor of approximately $10^5$. 
A set of novel formulas for bulk modulus, band gap, and oxygen evolution reaction activity are identified, which not only provide meaningful physical insights but also outperform previous formulas in accuracy and simplicity.}

\maketitle

\section{Introduction}\label{instruction}

The accurate prediction of physical properties is a key objective for many researchers. While deep learning methods, such as Graph Neural Networks (GNNs)\cite{gnn1,gnn2}, have demonstrated exceptional capabilities in predicting properties in materials and biological science\cite{CGCNN,alignn,biognn1,xue_accelerated_2016,lu_accelerated_2018}, they inherently operate as black boxes\cite{schmidt_recent_2019,rodriguez-perez_interpretation_2020}. Consequently, these methods cannot explain underlying physical mechanisms or provide insights, which significantly limits their value for fundamental scientific discovery\cite{mlexplain1,mlexplain2,mlexplain3}. Ideally, one would use an explicit formula to describe target properties and provide physical insights, rather than merely predicting numerical values\cite{ghiringhelli_big_2015,Reuter_2005}. However, due to the complexity of real-world systems, identifying such formulas remains a significant challenge.

Symbolic Regression (SR) method such as genetic programming \cite{gp}, SINDy \cite{SINDy}, and HI-SISSO \cite{HI-SISSO}, is a possible way to tackle the problem\cite{wang_symbolic_2019,koza_genetic_1994} and these methods have been used to model several physical properties\cite{wang_symbolic_2019,bartel_physical_2018,functional_form,ouyang_exploiting_2020,PhysRevMaterials.4.034204}. However, in the absence of experience and physical understanding of the data and physical process, one would use all input features and parameters to guarantee not missing the important ones. This turns the SR into a blind walking in a vast space. Consequently, it causes SR to merge physically irrelevant variables into the derived formulas. This results in equations that, while potentially accurate in fitting data, are physically incoherent and fail to reveal the true underlying mechanisms.

In recent years, Large Language Models (LLMs) have demonstrated potential in science, but still remain at an exploratory stage in many domains like material science\cite{boiko_autonomous_2023,Jablonka_2023,m_bran_augmenting_2024}. Many researchers have attempted to use LLMs as end-to-end symbolic regression engines, like Funsearch\cite{funsearch} and LLM-SR \cite{LLM-SR}, hoping to leverage the pre-trained scientific prior knowledge in LLM to directly extract physical frameworks from data. Although this approach has achieved some success in simple low-dimensional problems, the performance of LLMs in handling high-dimensional complex data is constrained by their fundamental nature as language models\cite{transformer_compos}. Lacking the intrinsic capability to process complex numerical patterns, they often fail to identify valid mathematical structures from such complex data\cite{math_cap_gpt}.

To overcome these limitations, we propose LangLaw, a language-model-guided framework designed to discover governing physical laws from scientific data. 
Our approach integrates the robust search capability of SR and the scientific knowledge and reasoning ability of LLMs, enabling the LLM to intelligently guide the SR process by identifying physically relevant variables and pruning the search space. 
We evaluated our framework on three representative materials property datasets: Bulk Modulus ($B_0$) for mechanical stability \cite{HI-SISSO}, Band Gap of lead-free perovskites for optoelectronic properties \cite{GAO2021150916}, and Oxygen Evolution Reaction (OER) activity for electrocatalytic performance \cite{OER}. 
It is worth noting that materials property data are often scarce due to experimental and computational challenges, which makes it more difficult to extract information using data-driven methods. The rich scientific knowledge embedded in LLMs provides critical assistance in overcoming this challenge, thereby enabling more effective formula discovery from limited data.

\section{Results}

\begin{figure}[h!]
    \centering
    \includegraphics[width=\linewidth]{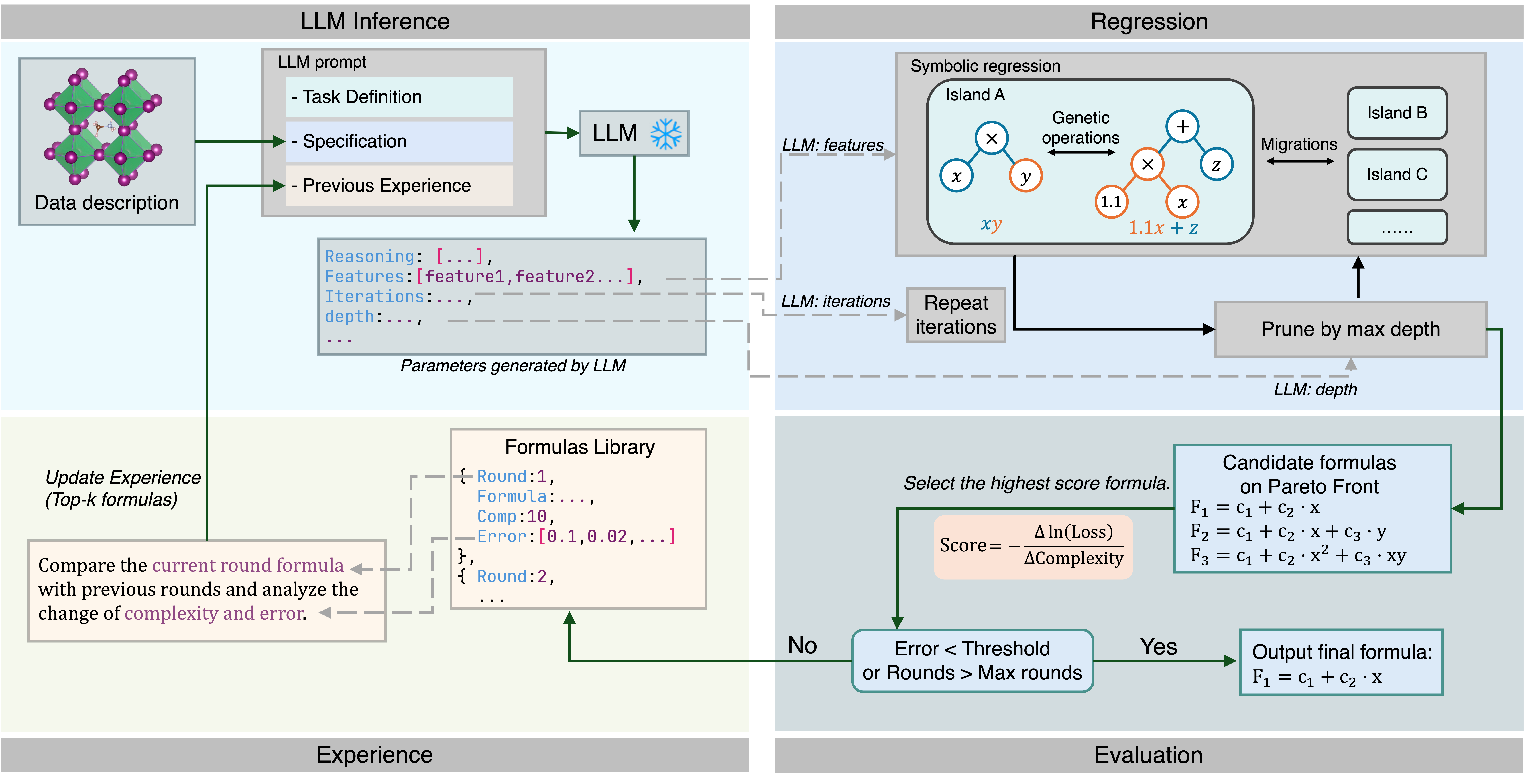}
    \caption{\textbf{Overview of the LangLaw framework.} The workflow is organized into four interconnected phases forming a closed loop: LLM inference (top left): The Large Language Model (LLM) acts as a reasoning agent. It analyzes the data description (e.g., crystal structure) and previous experience to generate search constraints, including feature selection, iteration counts, and tree depth. Regression (top right): These parameters act as control signals (purple dashed lines) to guide the Symbolic Regression (PySR) engine. The engine performs evolutionary searches using a parallel island model, evolving formula populations via genetic operations like crossover and mutation (details are shown in Supplementary Note S1). Evaluation (bottom right): Candidate formulas on the Pareto front are screened. The optimal formula is selected based on a score function that balances fitting loss and complexity\cite{gp}. Then if the error is lower enough or comes to the end round, the formula will be output, else the formula will be added into the Formulas Library. Experience (bottom left): The selected formulas and their performance metrics are stored in a formula library. This knowledge is formatted into prompts to update the LLM's experience, refining the search strategy for subsequent rounds.}
    \label{fig:method}
\end{figure}

The LangLaw framework operates as an iterative loop, as shown in Fig. \ref{fig:method}. First, the LLM analyzes the descriptions of input features (such as electron negativity, atomic radii, and ionization potential) to generate specific search parameters and choose features as inputs. 
This step mitigates combinatorial explosion and reduces the effective search space by a factor of approximately $10^5$ (see Supplemental Figure S8).

Here we used the Intern-S1~\cite{intern-s1}, which is a multimodal foundation model that enhances scientific reasoning capability, but other LLMs can also be used (see Supplementary Note S5). 
Guided by these instructions, the SR engine (implemented using PySR \cite{PySR}) performs a search to find candidate mathematical formulas. We record the results of each iteration, including the derived formulas, parameters, and fitting errors, into an Experience Pool. The LLM then reviews this historical data to identify effective variable combinations and refine the instructions for subsequent rounds. This feedback mechanism allows the system to progressively narrow down the search space and identify physically meaningful equations. 

In each step of the loop, the LLM analyzes the meaning of the input features and reviews the results from previous searches. 
Based on its scientific knowledge, the model suggests specific variables and parameters for the next search. It filters out features that do not make physical sense, even if they show a statistical correlation. The Symbolic Regression engine then performs the search using these constraints. Finally, it produces a set of formulas that balance high accuracy with low complexity.
Comprehensive details regarding the symbolic regression settings, the specific structure of LLM prompting, reasoning examples, and the implementation mechanism of the experience pool are elaborated in the Supplementary Note S2.

To evaluate the performance of LangLaw, we applied it to the task of modeling materials properties. 
We evaluated the discovered formulas based on prediction error and complexity. The formulas located on the Pareto front, as well as those identified in previous studies using other SR-based methods, are listed in Supplementary Table S4-S9. 

\subsection{Perovskite Bulk Modulus}

\begin{figure}[h!]
    \includegraphics[width=\linewidth]{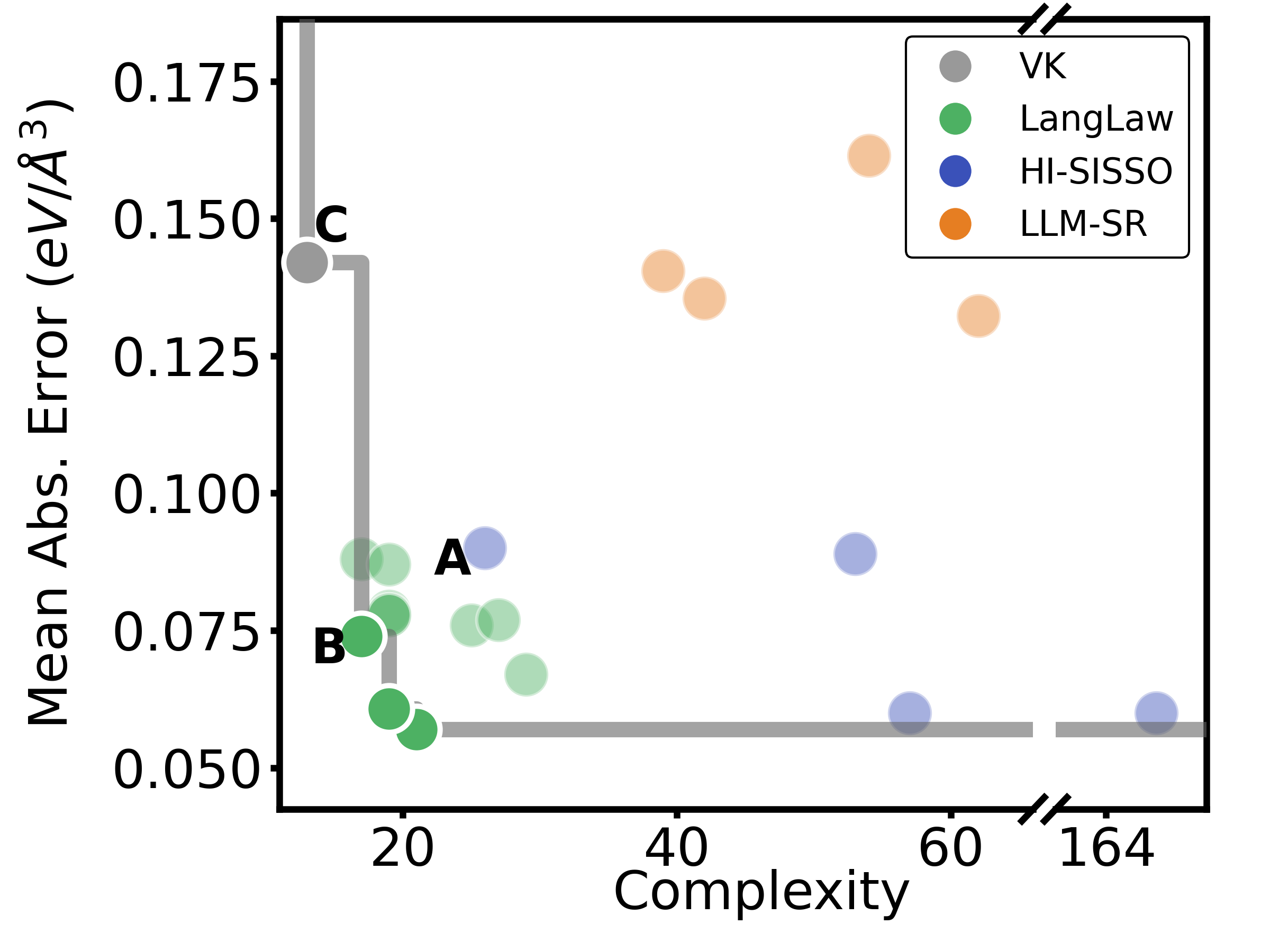}
    \caption{\label{fig:Pareto_front}\textbf{Performance comparison on the Perovskite Bulk Modulus dataset.} This plot shows the complexity and Mean Absolute Error of formulas of Bulk Modulus found by different methods: Verma and Kumar's formula (gray points), LangLaw (green points), HI-SISSO (blue points) and LLM-SR (yellow points). The gray line is the Pareto front. }
\end{figure}

The bulk modulus ($B_0$) measures a material's resistance to uniform compression and is a key indicator of mechanical stability. 
% For perovskites, which are widely used in photovoltaics, catalysis, and thermoelectrics, maintaining structural stability under operating conditions is essential~\cite{jonathan_perovskites,jena_halide_2019}. Furthermore, $B_0$ is a fundamental parameter in theoretical materials science, used for constructing equations of state, predicting phase transitions under pressure, and validating computational models such as Density Functional Theory (DFT). 
Identifying the factors that govern $B_0$ is critical for understanding intrinsic material properties and designing stable functional perovskites.
Previous research has attempted to predict $B_0$ using either empirical formulas or data-driven methods. Verma and Kumar proposed the following empirical relation\cite{VERMA2012210} (point C in Fig. \ref{fig:Pareto_front}):
\begin{equation}
    B_0^\text{VK} = C_0+C_1\frac{(n_A n_B)^{C_2}}{(a_0)^{3.5}},
    \label{eq:vk_formula}
\end{equation}
where $a_0$ is the lattice parameter, $n_A$ and $n_B$ are the oxidation number of the A/B-site ions, and $C_0, C_1, C_2$ are fitted constants. Alternatively, HI-SISSO~\cite{HI-SISSO} identified a formula from a set of 23 primary features, including valence orbital radius, electron affinity, ionization potential, and electronegativity (point B in Fig. \ref{fig:Pareto_front}):
\begin{equation}
    B_0^{\text{HI-SISSO}} = 2.99\frac{(IP_B-EA_B)^{0.419}(E_0)^{0.964}}{(a_0-5.09\times 10^{-4}\frac{EA_B n_A}{|r_{s,B}^{\text{cat}}-r_{s,B}|})^{2.75}},
    \label{eq:hisisso}
\end{equation}
where $IP_B$ and $EA_B$ denote the ionization potential and electron affinity of the B-site ion, respectively. $E_0$ represents the cohesive energy, while $r_{s,B}$ and $r_{s,B}^\text{cat}$ correspond to the radii of the valence $s$-orbital of the B-site ions in its neutral and cationic ($1+$) states respectively.
While Eq. (\ref{eq:hisisso}) offers improved accuracy compared to Eq. (\ref{eq:vk_formula}), its multiple coupled terms and fitted exponents compromise physical interpretability.

We utilized the same dataset and experimental settings as in HI-SISSO~\cite{HI-SISSO}. 
The results are presented in Fig. \ref{fig:Pareto_front}. LangLaw identified a set of formulas along the Pareto front, each keeping a balance between accuracy and simplicity. 
We selected the formula corresponding to point A for a detailed analysis:
\begin{equation}
    B_0^\text{LangLaw} = - \left(\frac{EA_{B}}{IP_{B}}\right) +  0.51 \left(\frac{n_{A} + 25.7}{a_0}- EN_{B} \right)  - 1.75
    \label{eq:ours}
\end{equation}
where the newly emerged $EN_B$ denotes the electron negativity of the B-site atom.

This formula provides a clear and interpretable linear relationship for the bulk modulus of perovskites: 

The first term $-(\frac{EA_B}{IP_B})$ can measure the "softness" of the electron cloud. A higher electron affinity $EA_B$ (stronger tendency to gain an electron) coupled with a lower ionization potential $IP_B$ (easier to lose an electron) indicates a more polarizable electron cloud. Ions with highly polarizable electron clouds are more easily deformed under stress, which reduces the resistance to compression and results in a lower bulk modulus. 
% The negative sign in front of the term indicates a larger $-(\frac{EA_B}{IP_B})$ value decreases $B_0$, consistent with the concept of ionic "softness". 

% This softness of electron cloud aligns with the Absolute Hardness Theory derived by Parr and Pearson\cite{parr_absolute_1983}, describing that the hardness of an atomic or ion can be formatted as $\eta \approx (IP-EA)/2$. They proved that this theory governs the interaction between Lewis acids and bases using Density Function Theory, where larger $\eta$ means higher hardness and harder for reaction.
% Our found first term defined as the softness of electron cloud extended the Absolute Hardness Theory from chemical reactivity to the interaction between perovskite's B-site and its bulk modulus, where the softer of the electron cloud of B-site ions decrease the mechanical resistance of lattices, serving as a bridge of SR to traditional calculation and theory derivation.

The second term can be divided into two parts. 
% $0.51 \left(\frac{n_{A} + 25.7}{a_0}- EN_{B} \right)$ contains two part.
The first part $\frac{n_{A} + 25.7}{a_0}$ resembles in form to the non-linear term $\frac{(n_A n_B)^{C_2}}{a_0^{3.5}}$ in the VK formula (Eq. \ref{eq:vk_formula}). In $ABO_3$ perovskites, charge neutrality imposes the constraint $n_A + n_B = 6$. 
% Consequently, the charge product $n_A n_B$ (ranging from 6 to 9) exhibits a strong positive correlation with the A-site oxidation state $n_A$. 
We also note that the lattice parameter $a_0$ varies within a narrow range (centered around 4 \AA). 
Therefore, our derived term can be regarded as an effective linear proxy for the non-linear term in the VK formula (Eq.~\ref{eq:vk_formula}). 
The $-EN_B$ part performs an ionic correction. Higher $EN_B$ weakens the $B-O$ ionic bonds, which softens the lattice leading to lower bulk modulus.

% Noteworthy, Eq.\ref{eq:ours} generalizes Pearson’s Absolute Hardness Theory, which is hard to quantify and limited to molecular chemistry\cite{parr_absolute_1983}. By quantifying the cloud softness term $-(\frac{EA_{B}}{IP_{B}})$ and integrating it with structural factors, we formalize this concept into a precise physical law bridging electronic states and mechanical stiffness.

\begin{figure}%[h]
    \includegraphics[width=\linewidth]{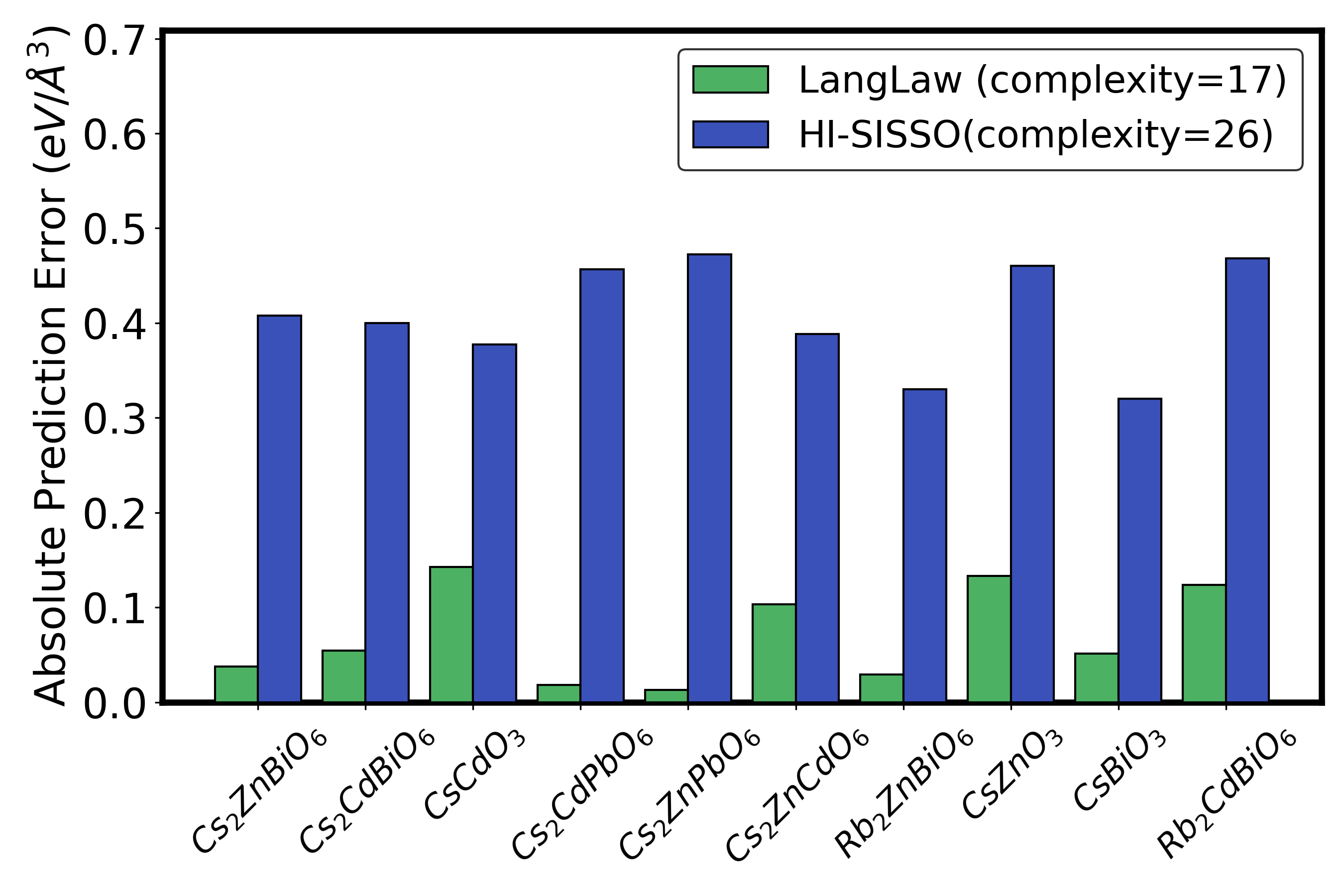}
    \caption{\label{fig:OOD_test}\textbf{Evaluation of out of distribution generalization capability.} The bar chart compares the absolute prediction error of our discovered linear formula against the high-complexity HI-SISSO model on 10 screened perovskite materials not seen during training. Our method (green bars) consistently yields lower prediction errors across diverse compositions compared to HI-SISSO (blue bars), demonstrating superior transferability and robustness in data-scarce scenarios.}
\end{figure}

To evaluate the transferability of our model to out-of-distribution (OOD) data, we tested Eq. \ref{eq:ours} on 10 perovskites selected from 7,308 single ($ABO_3$) and double ($A_2BB'O_6$) structures from the same study\cite{HI-SISSO}. These specific materials were chosen as OOD because their bulk modulus values and the double perovskite structure are rare in the training set. 
% This process also simulated a material discovery scenario where data is scarce. 
As shown in Fig. \ref{fig:OOD_test}, our linear formula achieves remarkably lower prediction errors than Eq. \ref{eq:hisisso}, demonstrating superior generalization to new material compositions.

\subsection{Double Perovskite Band Gap}
\begin{figure}%[h]
    \includegraphics[width=0.9\linewidth]{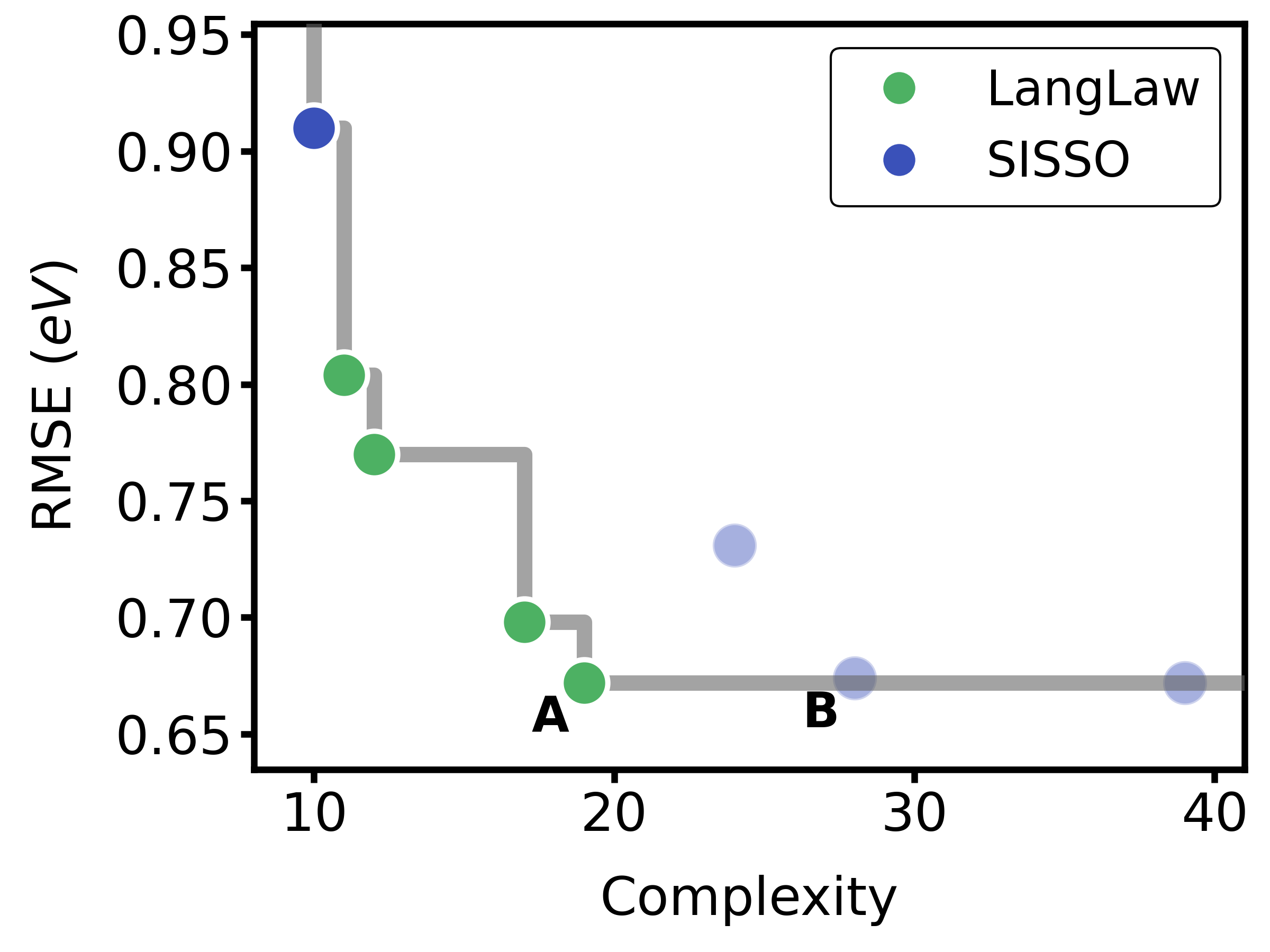}
    \caption{\label{fig:Pareto_front_bandgap}\textbf{Performance comparison on the Band Gap dataset.} This plot shows the complexity and Mean Absolute Error of formulas of Band Gap found by LangLaw (green points) and SISSO (blue points). The gray line is the Pareto front. }
\end{figure}

Next, we applied LangLaw to the band gap prediction problem of lead-free double perovskites ($A_2BB'X_6$). Band gap is a core parameter for screening photovoltaic materials\cite{JRC122879}. 
% For solar cell applications, a suitable material band gap (typically 1.1-1.7 eV) is a prerequisite for achieving efficient solar spectrum absorption and high open-circuit voltage\cite{JRC122879}. 
We employed a dataset containing 745 $A_2BB'X_6$ materials and their band gaps extracted from Ref. \cite{GAO2021150916}, aiming to discover key physicochemical factors affecting the band gap. The results are shown in Fig.~\ref{fig:Pareto_front_bandgap}.

Among the candidates found, we identified a highly interpretable formula that balances simplicity and accuracy (point A in Fig. \ref{fig:Pareto_front_bandgap}):
\begin{equation}
    E_g^\text{LangLaw} =0.056\left(\frac{X_X^3}{V_B^4}\right)+\frac{2.66}{R_XV_AX_{B'}^2},
    \label{eq:our_formula_bandgap}
\end{equation}
where $V_A$ and $V_B$ are the valence electrons of the A- and B-site atoms, respectively, $R_X$ is the ionic radius of the X-site anion, and $X_X$ and $X_{B'}$ are the electronegativity of the X- and B'- site cation, respectively. 

We compare Eq.~\ref{eq:our_formula_bandgap} with the formula derived by the SISSO method~\cite{SISSO,baloch2022bandgap} (point B in Fig. \ref{fig:Pareto_front_bandgap}) on the same dataset, given in Eq.~(\ref{hisosso_bandgap}):
\begin{equation}
\begin{split}
    E_g^\text{SISSO} &= 0.573 + 0.053\left(\frac{X_X^3}{V_B^4}\right) \\
    &+ 3.509\left(\frac{1}{X_{B'}^2 \sqrt{V_A}}\right) - 0.154\left(\frac{\sqrt{Z_X}}{R_B}\right),
\end{split}
\label{hisosso_bandgap}
\end{equation}
where $Z_X$ is atomic number of the X-site anion and $R_B$ is the ionic radii of the B-site cation. 

Interestingly, both formulas incorporate the term $\frac{X_X^3}{V_B^4}$ with nearly identical coefficients. 
% as well as the inverse square of the B$'$-site electronegativity ($1/X_{B'}^2$). 
This suggests that these factors play a major role in determining the band gap. 
% The second terms initially appear distinct. To understand their relationship, we analyzed the statistical distribution of the variables in the training dataset. We reveals that replacing the variable $R_X$ (ionic radius of the X-site anion) with its mean value results in a negligible increase in prediction error, suggesting that $R_X$ functions primarily as a normalization constant rather than a driving variable in this context. The second term mathematically converges to a dependence proportional to $X_{B'}^{-2}$. This finding aligns with the SISSO formula's $1/X_{B'}^2$ factor. 
There is also a pair of similar terms, both containing $X_{B'}$ and $V_A$ in the denominator. The difference lies in that our formula includes an extra term $R_X$. We analyzed the distribution of $R_X$ in the dataset and found that its variation is relatively small. By replacing $R_X$ with the mean value of $R_X$ in the dataset, we observed only a slight increase in prediction error. Furthermore, we noted that $V_A$ in the two formulas differs by a square root. Numerical analysis reveals that this term takes only the values 1 or 11, meaning the difference introduced by the square root can also be treated as a constant. Consequently, this term in our two formulas becomes highly similar as well. 
However, the LangLaw formula (Eq.~\ref{eq:our_formula_bandgap}) has advantages in conciseness. 

% Our formula (Eq.~\ref{eq:our_formula_bandgap}) shows functional similarities to the baseline formula. Both formulas include terms proportional to the inverse fourth power of the B-site valence ($1/V_B^4$) and the inverse square of the B$'$-site electronegativity ($1/X_{B'}^2$). 
% The main difference lies in the dependence on anion properties: Eq.~\ref{eq:hisosso_bandgap} relies solely on cation information ($R_B$, $V_B$, $X_{B'}$, and $V_A$), while our formula (Eq.~\ref{eq:our_formula_bandgap}) includes the electronegativity ($X_X$) and radius ($R_X$) of the anion.
% This indicates that in our model, the band gap is described by the properties of the anion p-orbitals in addition to the cationic framework.
% \subsection{OER Activities}

\subsection{OER Activity}
Finally, we applied LangLaw to search for new formulas for the oxygen evolution reaction (OER) activity of oxide perovskites. The OER is a key rate-limiting step that determining the overall efficiency in electrocatalysis\cite{OER1,OER2}. Oxide perovskites are considered highly promising candidates for OER catalysts due to their rich compositional tunability, flexible electronic structure, and excellent chemical stability\cite{yin_oxide_2019}. 

\begin{figure}%[h]
    \includegraphics[width=0.9\linewidth]{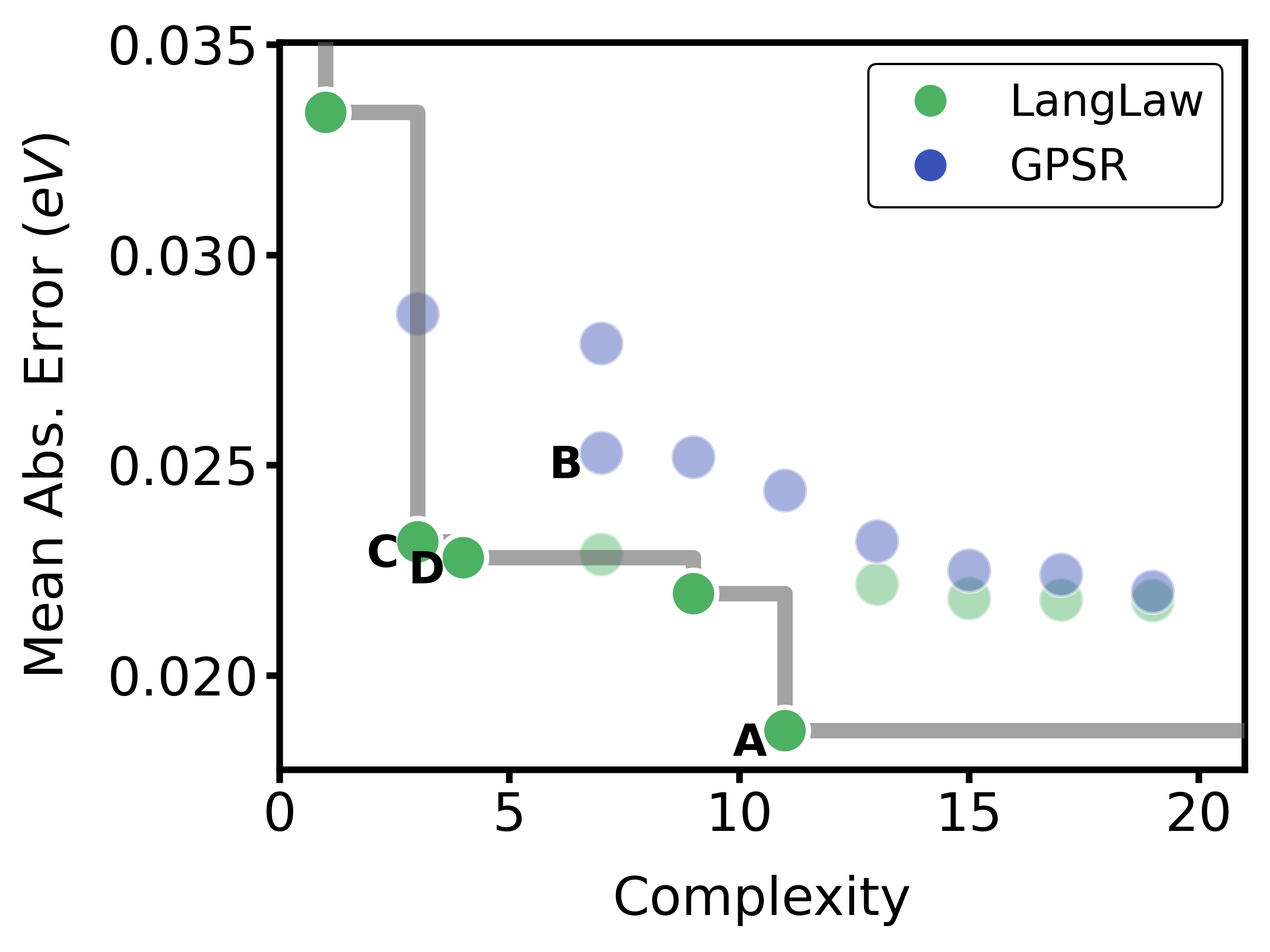}
    \caption{\label{fig:Pareto_front_oer}\textbf{Performance comparison on the OER activity dataset.} This plot shows the complexity and Mean Absolute Error of formulas of OER activity found by LangLaw (green points) and GPSR (blue points). The gray line is the Pareto front. }
\end{figure}

A previous study~\cite{OER} proposed GPSR (symbolic regression with genetic programming) and derived a formula (point B in Fig. \ref{fig:Pareto_front_oer}) to predict OER activity, measured by the potential versus the reversible hydrogen electrode ($V_\text{RHE}$):
\begin{equation}
    V_\text{RHE}^\text{GPSR}=1.554\frac{\mu}{t}+1.092,
    \label{eq:gpsr}
\end{equation}
where $\mu$ is the octahedral factor that reflects the local geometry of the $BO_6$ octahedron and $t$ is the tolerance factor that describes the global distortion of the crystal structure.
This model was trained on 18 synthesized ABO$_3$ perovskites, all measured under unified and comparable experimental conditions, ensuring data reliability and consistency. The formula was subsequently tested on 5 high-activity, stable perovskites screened from 3,545 perovskites. 

Using the same OER activity dataset, we identified a formula with higher accuracy (point A in Fig. \ref{fig:Pareto_front_oer}):
\begin{equation}
    V_\text{RHE}^\text{LangLaw}=(\mu + 0.127)\times (3.24+\frac{0.0016}{t-1.1}).
    \label{eq:our_oer}
\end{equation}

Our formula (Eq.~\ref{eq:our_oer}) also relates the OER activity to two geometric factors $\mu$ and $t$. 
We note that the term involving $t$ has a very small coefficient of 0.0016, suggesting that $t$ likely has a limited influence on the outcome. Numerical analysis shows that the value of this term is only around 0.1, which is considerably smaller than the constant 3.24 added to it. To validate this assumption, we also identified some formulas containing only $\mu$, such as those corresponding to points C ($3.810^\mu$) and D ($\log(\mu)+2.623$) in Fig. \ref{fig:Pareto_front_oer}, whose prediction errors remain lower than that of the GPSR formula (Eq. \ref{eq:gpsr}).
% Unlike the previous linear ratio (Eq.~\ref{eq:gpsr}), our formula suggests that the local geometric distortion $\mu$ modulates the effect of the global lattice tolerance $t$ in a non-linear way. Numerical analysis of the dataset reveals that the tolerance factor $t$ is concentrated around 1.1. In this regime, the term $\frac{0.0016}{t-1.1}$ yields a relatively small value (approximately 0.1). When comparing this to the GPSR model (Eq.~\ref{eq:gpsr}), where $t$ appears simply as a linear denominator, the variation of $1/t$ is minimal due to the narrow distribution of $t$, resulting in a negligible modulation of the octahedral factor $\mu$. Our formula employs a non-linear term that, despite its small magnitude, provides a sensitive correction. This multiplicative coupling provides a more precise description of the specific structural constraints and improves fitting accuracy.

Using the three datasets, bulk modulus, band gap, and OER activity, we also compared LangLaw with other methods, as shown in Table~\ref{tab:All_experiments_comparison}. 
LLM-SR~\cite{LLM-SR} is a recently proposed approach for scientific equation discovery that leverages the scientific knowledge and code generation capabilities of LLMs. 
The formulas identified by LLM-SR not only exhibit greater complexity but also result in higher predictive errors. Deep learning methods, including CGCNN~\cite{CGCNN} and ALIGNN~\cite{alignn}, typically require a large amount of data and tend to overfit in small-data scenarios. 
For bulk modulus prediction, LangLaw outperforms both CGCNN and ALIGNN. 
Notably, on OOD data, LangLaw achieves the lowest RMSE of 0.0851, which is half that of ALIGNN (0.167) and five times lower than CGCNN (0.401).
For band gap prediction, CGCNN and ALIGNN also yield higher predictive errors. We did not apply CGCNN or ALIGNN to the OER activity dataset, as it consists of only 18 data points. 
This comparison highlights the significant advantage of LangLaw in extracting robust and transferable physical laws from small datasets, where pure data-driven methods often struggle.

\section{Conclusion}
In conclusion, we introduced a Language-guided SR (LangLaw) framework that combines LLMs with SR to discover interpretable physical laws from high-dimensional data. By leveraging the scientific knowledge and reasoning capabilities of LLMs to guide the regression process, our method overcomes the key limitation of traditional SR that tends to produce complex, unphysical expressions, and efficiently uncovers formulas that are both accurate and insightful. 
% We validated the framework on three tasks of materials property prediction. Notably, for the perovskite bulk modulus, we discovered a simple linear law that highlights the critical role of B-site covalency in mechanical stability. 

Beyond advancing symbolic regression, this work significantly extends the role of LLMs in materials design. Rather than serving merely as predictors or text generators, LLMs here act as knowledge-guided search engines, directly shaping the discovery of fundamental physical relationships. This demonstrates a novel pathway for LLMs to contribute to interpretable, mechanism-driven materials design, offering researchers a practical and principled tool to extract governing scientific laws from complex, real-world data.

\section{Methods}

\subsection{Implementation}

We implement the overall workflow of the L‑SR framework based on the PySR library~\cite{PySR}. PySR employs a multi‑island genetic programming algorithm, representing mathematical expressions as tree structures that evolve through tournament selection, crossover, and mutation within isolated populations. In each round, the LLM analyzes the dataset, textual descriptions, and an experience pool that stores the selected feature sets, hyperparameters, and fitting errors from previous search rounds. Based on this information, the LLM identifies relevant physical variables and provides the feature subset, the maximum tree depth, and the number of evolutionary iterations to be used in the current round of SR search. This design restricts the search space to physically relevant variables and avoids the generation of overly complex functional forms. During the search, continuous constants are optimized using gradient-based methods, and the algorithm identifies a Pareto front of formulas to balance accuracy and complexity. The output formula from each round of SR is evaluated using the scoring function described in Supplementary Note S4. The details of the PySR and LLM implementations are presented in Supplementary Notes S1-S2. For the selection of LLMs, we mainly used Intern-S1~\cite{intern-s1}, but other LLMs are also applicable. The LLM settings and comparisons are detailed in Supplementary Note S5.

% For the SR method we used the PySR library, which uses a multi-island genetic programming algorithm. In this library, mathematical expressions are represented as tree structures that evolve through tournament selection, crossover, and mutation within isolated populations. The selected features input constrains the search space and the parameters set by the LLM control the complexity and duration of the evolution. During the search, continuous constants are optimized using gradient-based methods, and the algorithm identifies a Pareto front of formulas to balance accuracy and complexity. The final output of each rounds SR is evaluated by a score function described in Supplementary Note S4 and the detailed workflow of PySR can be found in Supplementary Note S1.

% To evaluate L-SR, we compared it with three categories of baselines: traditional SR methods (HI-SISSO, SISSO, and GPSR), recent LLM-based SR method (LLM-SR), and deep learning methods (CGCNN and ALIGNN). The parameters settings are provided below and detailed in Supplementary Notes S6-S7. 
% We compared with GNN using CGCNN and ALIGNN to evaluate the precision and generalization of L-SR, LLM based SR method using LLM-SR to evaluate the performance of L-SR with relative method and tradisional SR method: HI-SISSO, SISSO and GPSR to demostrate the effectiveness of LLM guidance. The details about GNN and LLM-SR are shown in Supplementary Note S6-S7. The parameters of baslines and L-SR are described below.

\subsection{Experimental settings}

To evaluate LangLaw, we compared it with three categories of baselines: traditional SR methods (HI-SISSO, SISSO, and GPSR), recent LLM-based SR method (LLM-SR), and deep learning methods (CGCNN and ALIGNN). The parameters settings are provided below and detailed in Supplementary Notes S6-S7. 

The results reported for HI-SISSO, SISSO, and GPSR in Table~\ref{tab:All_experiments_comparison} were calculated using the formulas in their respective original papers. 
For LLM-SR, we utilized the code released with the paper, adapting the prompts for our experimental context. In line with the LLM-SR paper, we employed GPT-4o as the LLM backbone and used the temperature of 0.8 for LLM inference. 
For CGCNN and ALIGNN, we adopted their publicly released implementations. CGCNN is trained using a hidden dimensions of 128, a batch size of 256, three message passing layers, a learning rate of 1e-2, a radius cutoff of 8.0 \AA, the nearest neighbors setting to 12, and is trained from scratch for 2000 epochs. ALIGNN follows its original setup with a learning rate of 1e-3, a batch size of 32, and is trained from scratch for 2000 epochs. 

% For ALIGNN, we follow the original implementation details of ALIGNN released in the paper using the publicly available code from the authors to train and evaluate it on the three main experiments dataset. For the hyper-parameter settings we set the learning rate to 1e-3, batch size to 32 , and training it for 2000 epochs.

% For LLM-SR, we implement the public code released with the paper with our specialized prompts to run the LLM-SR framework. For the LLM selection we followed the usage of GPT-4o described in the LLM-SR paper to guarantee the performance. For the hyper-parameters to run we used the temperature of 0.8 for LLM inference as the same as the paper.

% For the HI-SISSO, SISSO and GPSR outcome in Table .\ref{tab:All_experiments_comparison}, we followed the original papers' instruction to reproduce the same outcome and evaluate the performance and errors.

% \subsection{L-SR}

In the PySR component of LangLaw, we set the populations to 31, the population\_size to 27 and the max\_size to 30. The L1 distance is selected as the loss function. The allowed operator set comprises $+$, $-$, $\times$, $\div$, and exponentiation \textasciicircum{}, with constrain parameter \{-1, 1\} of the pow operator. 
% For model selection, we set it to "best" in three main experiments to find the most interpretable formula and to "accuracy" in ablation to precisely quantify L-SR method's ability to discover formulas.
For the LLMs module, the temperature is set to 1.0 for balancing performance. The experience pool retains the top-$k$ historical formulas with $k=15$. Example prompts are provided in Supplementary Note S2. The search ends when the error falls below a threshold of 0.001 or after a maximum of 100 rounds. 
% For the Experience part in L-SR we select top-$K$ previous formulas where k is set to 15 for sufficient experience for refinement. The example of prompts used in L-SR is described in Appendix. The error threshold is set to 0.001 and the max rounds is set to 100 in three main experiments. 

\section*{Data availability}
ALL datasets used in this study are publicly available (see Supplementary Note S3).

\section*{Code availability}
The code for reproducing the findings in this work is available at \url{https://github.com/nakonako4/langlaw}. 

\section*{Acknowledgments}
This work was supported by New Generation Artificial Intelligence-National Science and Technology Major Project(2025ZD0121802), Shanghai Committee of Science and Technology, China (Grant No. 23QD1400900), National Key Research and Development Program of China (Grant No. 2025YFC2311702 and 2025YFC2311703), the National Natural Science Foundation of China (Grant No. 12404291 and 32371299) and the Innovation Program for Quantum Science and Technology (Grant No. 2024ZD0300102). Y.G. did this work during his internship at Shanghai Artificial Intelligence Laboratory. 

\section*{Author contributions}
M.S. conceived the idea and led the research. Y.G. developed the code and performed the experiments. W.Y., J.L., and M.S. analyzed the results. C.L., D.Z., and L.B. contributed technical ideas. Y.G. and M.S. wrote the first draft. All authors discussed the results and reviewed the manuscript. 

\section*{Competing interests}
The authors declare no competing interests.

%%===========================================================================================%%
%% If you are submitting to one of the Nature Portfolio journals, using the eJP submission   %%
%% system, please include the references within the manuscript file itself. You may do this  %%
%% by copying the reference list from your .bbl file, paste it into the main manuscript .tex %%
%% file, and delete the associated \verb+\bibliography+ commands.                            %%
%%===========================================================================================%%
\enlargethispage{1\baselineskip}
\bibliography{sn-bib}% common bib file
%% if required, the content of .bbl file can be included here once bbl is generated
%%\input sn-article.bbl
\newpage
\section*{Figures}

\begin{sidewaystable}[h!]
\caption{\textbf{Performance comparison on three material science tasks.} This table reports the formula complexity and prediction errors for bulk modulus, band gap, and oxygen evolution reaction ($V_\text{RHE}$). The selection of MAE or RMSE as the evaluation metric for each column follows the convention used in the respective source studies. HI-SISSO serves as the symbolic regression baseline for Bulk Modulus, while SISSO is the specific baseline for the Band Gap task and GPSR is for the OER task. These methods are compared alongside the LLM-based LLM-SR and deep learning models CGCNN and ALIGNN. ID refers to the test error on the in-distribution dataset, and OOD refers to the generalization error on out-of-distribution samples. Bold numbers indicate the best results.}\label{tab:All_experiments_comparison}
\begin{tabular*}{\textheight}{@{\extracolsep\fill}lcccccccc}
\toprule%
&\multicolumn{3}{c}{Bulk Modulus}&\multicolumn{2}{c}{Band Gap}&\multicolumn{3}{c}{$V_\text{RHE}$}\\
Methods & Complexity$\downarrow$ & \makecell{ID\\ MAE($\text{eV/}$\AA$^3$)$\downarrow$} & \makecell{OOD\\ RMSE($\text{eV/}$\AA$^3$)$\downarrow$} & Complexity$\downarrow$ & RMSE($\text{eV}$)$\downarrow$& Complexity$\downarrow$& \makecell{ID\\ MAE($\text{eV}$)$\downarrow$} & \makecell{OOD\\ MAE($\text{eV}$)$\downarrow$} \\
\midrule
HI-SISSO\cite{HI-SISSO} & 26 & 0.090 & 0.411 & --- & --- & --- & --- & --- \\
SISSO\cite{SISSO}     & --- & --- & --- & 39 & 0.672 & --- & --- & --- \\
GPSR\cite{OER}     & --- & --- & --- & --- & --- & 7 & 0.0253 & \textbf{0.0209} \\

LLM-SR\cite{LLM-SR}   & 63 & 0.140  & 3.93 & 70 & 0.669 & 54  &  0.0147 & 0.108 \\
CGCNN\cite{CGCNN}    &  ---  & 0.0903 & 0.401 &  --- & 1.053&  --- & ---  & --- \\
ALIGNN\cite{alignn}   &  ---  & 0.0784 & 0.167 &  --- & 1.114&  --- & ---    & --- \\
LangLaw & \textbf{17}& \textbf{0.0739} & \textbf{0.0851} & \textbf{19}& \textbf{0.672}& 11 & \textbf{0.0187}  &0.0225 \\
\botrule
\end{tabular*}
\end{sidewaystable}

\end{document}